\documentclass[aps,prev,twocolumn,preprintnumbers,floatfix,nofootinbib]{revtex4-1}
\usepackage{graphicx}
\usepackage{epstopdf}
\usepackage{mathrsfs}
\usepackage{amssymb}
\usepackage{verbatim}
\usepackage{color}
\usepackage{multirow}

\usepackage{amsmath}
\usepackage{epsfig}
\usepackage{ulem}
\usepackage{graphics}
\usepackage{indentfirst}

\newcommand{\pb}{\ensuremath{\bar{\Phi}}}
\newcommand{\xb}{\ensuremath{\bar{X}}}

\newcommand{\te}{\ensuremath{\theta}}

\newcounter{RomanNumber}
\newcommand{\MyRoman}[1]{\setcounter{RomanNumber}{#1}\Roman{RomanNumber}}

\def\stacksymbols #1#2#3#4{\def\theguybelow{#2}
    \def\vp{\lower#3pt}
    \def\sp{\baselineskip0pt\lineskip#4pt}
    \mathrel{\mathpalette\intermediary#1}}

\def\intermediary#1#2{\vp\vbox{\sp
     \everycr={}\tabskip0pt
     \halign{$\mathsurround0pt#1\hfil##\hfil$\crcr#2\crcr
              \theguybelow\crcr}}}

\def\be{\begin{equation}}
\def\ee{\end{equation}}
\def\bea{\begin{eqnarray}}
\def\eea{\end{eqnarray}}

\def\sp{\;\;\;,\;\;\;}

\def\lsim{\raise0.3ex\hbox{$\;<$\kern-0.75em\raise-1.1ex\hbox{$\sim\;$}}}
\def\gsim{\raise0.3ex\hbox{$\;>$\kern-0.75em\raise-1.1ex\hbox{$\sim\;$}}}

\def\inbar{\,\vrule height1.5ex width.4pt depth0pt}

\def\IC{\relax\hbox{$\inbar\kern-.3em{\rm C}$}}
\def\IQ{\relax\hbox{$\inbar\kern-.3em{\rm Q}$}}
\def\IR{\relax{\rm I\kern-.18em R}}
 \font\cmss=cmss10 \font\cmsss=cmss10 at 7pt
\def\IZ{\relax\ifmmode\mathchoice
 {\hbox{\cmss Z\kern-.4em Z}}{\hbox{\cmss Z\kern-.4em Z}}
 {\lower.9pt\hbox{\cmsss Z\kern-.4em Z}}
 {\lower1.2pt\hbox{\cmsss Z\kern-.4em Z}}\else{\cmss Z\kern-.4em Z}\fi}

\def\comment#1{}

\def\u1x{U(1)_X}
\newcommand{\nc}{\newcommand}
\nc{\LL}{L}
\nc{\vv}{\tilde{v}}
\nc{\ccdot}{\!\cdot\!}
\nc{\gsm}{G_{SM}}
\nc{\vfive}{\mathbf{5}\oplus\mathbf{\overline{5}}}
\nc{\vten}{\mathbf{10}\oplus\mathbf{\overline{10}}}
\nc{\zhol}{Z^{\rm hol}}
\nc{\xfb}{\,{\rm fb}}

\begin{document}

%
%

\preprint{ACT-10-14, MIFPA-14-26}

\vspace*{1mm}

\title{Helical Phase Inflation}

\author{Tianjun Li$^{a}$}
\email{tli@itp.ac.cn}
\author{Zhijin Li$^{b}$}
\email{lizhijin@physics.tamu.edu}
\author{Dimitri V. Nanopoulos$^{b, c}$}
\email{dimitri@physics.tamu.edu}

\vspace{0.1cm}
\affiliation{
${}^a$ State Key Laboratory of Theoretical Physics
and Kavli Institute for Theoretical Physics China (KITPC),
      Institute of Theoretical Physics, Chinese Academy of Sciences,
Beijing 100190, P. R. China; \\
School of Physical Electronics,
University of Electronic Science and Technology of China,
Chengdu 610054, P. R. China
 }
\affiliation{
${}^b$
George P. and Cynthia W. Mitchell Institute for
Fundamental Physics and Astronomy,
Texas A\&M University, College Station, TX 77843, USA}

 \affiliation{
${}^c$
Astroparticle Physics Group, Houston Advanced Research Center (HARC), Mitchell Campus, Woodlands, TX 77381, USA;\\
Academy of Athens, Division of Natural Sciences,
28 Panepistimiou Avenue, Athens 10679, Greece}

\begin{abstract}

We show that the quadratic inflation can be realized by the phase of a complex field with helicoid potential. Remarkably, this helicoid potential can be simply realized in minimal supergravity. The global $U(1)$ symmetry of the K\"ahler potential introduces a flat direction and evades the $\eta$ problem automatically. So such inflation is technically natural. The phase excursion is super-Planckian as required by the Lyth bound, while the norm of the complex field can be suppressed in the sub-Planckian region. This model resolves the ultraviolet sensitive problem of the large field inflation,
besides, it also provides a new type of monodromy inflation in supersymmetric field theory with consistent field stabilization.
\end{abstract}

\maketitle


\setcounter{equation}{0}



\subsection*{Introduction}
Inflation \cite{Guth:1980zm} as a model of the early Universe plays a crucial role in modern comology. It beautifully solves the horizon, flatness, and monopole problems, as well as explains the density fluctuation observed in the cosmic microwave background. Some details on the inflationary process are obtained from recent observations of
the Planck \cite{Ade:2013uln} and BICEP2 \cite{Ade:2014xna} experiments. It shows the inflation scale is about $10^{16}$ GeV, close to the scale for Grand Unified Theory (GUT). To generate slow-roll inflation, the scalar field $\phi$ should have sufficiently flat potential $V(\phi)$ so that its mass is hierarchically smaller than the Hubble constant
\begin{equation}
\eta\equiv M_P^2\frac{V''}{V}\simeq \frac{m_\phi^2}{3H^2}\ll1, \label{eta}
\end{equation}
where $M_P$ is the reduced Planck scale.
At the classical level, the potential can be set sufficiently flat by hand. However, the inflaton as a scalar field receives dangerous quantum corrections and even serious quantum gravity corrections if there is super-Planckian field excursion. The crucial challenge for a sensible inflation model is to protect the flat condition against these dangerous corrections.

At the GUT scale physics is considered to be supersymmetrical and the quantum corrections on the inflaton potential are effectively suppressed by supersymmetry \cite{ENOT}. However, the flatness of the potential is significantly changed in supergravity. The F-term scalar potential is proportional to a factor $e^K$, $K$ is the K\"ahler potential and contains a term $\Phi\pb$ in minimal supergravity. The factor $e^{\Phi\pb}$ generates an inflaton mass close to the Hubble scale and
 hence breaks the slow-roll condition (\ref{eta}). The $\eta$ problem is absent in no-scale supergravity \cite{Ellis:2013xoa}, in which the K\"ahler potential is initially designed to solve the cosmological flatness problem \cite{Cremmer:1983bf}. Alternatively, one can introduce a shift symmetry $\Phi\rightarrow\Phi+iC$ \cite{Kawasaki:2000yn} in the K\"ahler potential so that $e^K$ is flat along the shift direction$^1$. \footnotetext[1]{The shift symmetry can be slightly broken to get inflationary models with a broad range of tensor-to-scalar ratio $r$ \cite{Li:2013nfa}.}

For single field slow-roll inflation, the Lyth bound \cite{Lyth:1996im} indicates a super-Planckian inflaton excursion $\Delta\phi\sim 10 \, M_P$ for large tensor modes, which makes the effective theory description of inflation problematic. In the Wilsonian sense, there are higher dimensional operators from quantum gravity effects that are suppressed by the Planck mass $M_P$ and irrelevant in the sub-Planckian region. However, once the inflaton becomes super-Planckian, inflation is sensitive to the higher dimensional operators and the theory is not reliable unless it is
Ultraviolet (UV)-completed \cite{Baumann:2014nda}.

Problems from quantum gravity corrections can be avoided if the super-Planckian field excursion is effectively realized in the sub-Planckian region.
Considering the phase of a complex scalar field, or the pseudo-Nambu-Goldstone boson (PNGB) in gauge symmetry breaking scenario \cite{Freese:1990rb,Kim:2004rp,Baumann:2010nu,McDonald:2014oza}, the phase can have super-Planckian displacement while the magnitude of complex field remains sub-Planckian. Besides, the combination of multi sub-Planckian fields may lead to effective super-Planckian excursion \cite{Liddle:1998jc, Ashoorioon:2009wa}. Another attractive and widely studied model is the monodromy inflation \cite{McAllister:2008hb, Kaloper:2008fb}, in which the inflaton is an axion obtained from string compactification and evolves periodically while all the factors except the potential remain the same.

In this letter, we present a new inflation model with helicoid potential. This potential is designed to realize super-Planckian inflaton excursion with sub-Planckian fields and the inflation is driven by the phase of a complex field, so that we can keep away from dangerous quantum gravity corrections. Remarkably, the helicoid scalar potential can be simply obtained in minimal supergravity, and the well-known $\eta$ problem is automatically solved without any extra symmetry. The phase inflation also leads to a new type of monodromy in supersymmetric field theory with strong field stabilization.

\subsection*{Helicoid Potential}

Now we give the supergravity realization of the helicoid potential in the simplest case.
We consider two chiral superfields $\Phi$ and $X$ in minimal supergravity, the K\"ahler potential is
\begin{equation}
K=\Phi\pb+X\xb-g(X\xb)^2, \label{ka}
\end{equation}
where the higher order term $g(X\xb)^2$ is introduced to stabilize the field $X$ at $X=0$ \cite{Ellis:1984bs, Kallosh:2010xz}.
Besides, we use the following superpotential
\begin{equation}
W=a\frac{X}{\Phi}\ln\Phi. \label{sp}
\end{equation}
The superpotential is singular at $\Phi=0$ with monodromy\footnote[2]{An interesting proposal based on multivalueness of complex function with fractional power is studied in \cite{Harigaya:2014eta}.}
\begin{equation}
\Phi\rightarrow \Phi e^{2\pi i}, ~~W\rightarrow W+2\pi ai\frac{X}{\Phi}. \label{mono}
\end{equation}
In field theory, singularity appears when a massless field is integrated out. An explicitly realization of this monodromy will be provided based on supersymmetric field theory in next section.

It is obvious that the K\"ahler potential preserves the global $U(1)$ symmetry
for $\Phi$, which is broken by the superpotential. Thus, our model is
technically natural since there is a global $U(1)$ symmetry in the $a=0$ limit~\cite{tHooft}.

The F-term scalar potential is determined by the K\"ahler potential and superpotential as follows
\begin{equation}
V=e^K(K^{i\bar{j}}D_iW D_{\bar{j}}\bar{W}-3W\bar{W}).
\end{equation}
As the field $X$ is stabilized at $X=0$, the above potential is significantly simplified as below
\begin{equation}
\begin{split}
V&=e^{\Phi\pb}W_X\bar{W}_{\bar{X}} \\
&=a^2e^{r^2}\frac{1}{r^2}((\ln r)^2+\te^2),
\end{split}\label{po}
\end{equation}
where $\Phi \equiv re^{i\te}$. The quadratic term $\theta^2$ appears in the potential because of
the monodromy (\ref{mono})
with respect to the origin.

\begin{figure}
\centering
\includegraphics[width=80mm, height=70mm,angle=0]{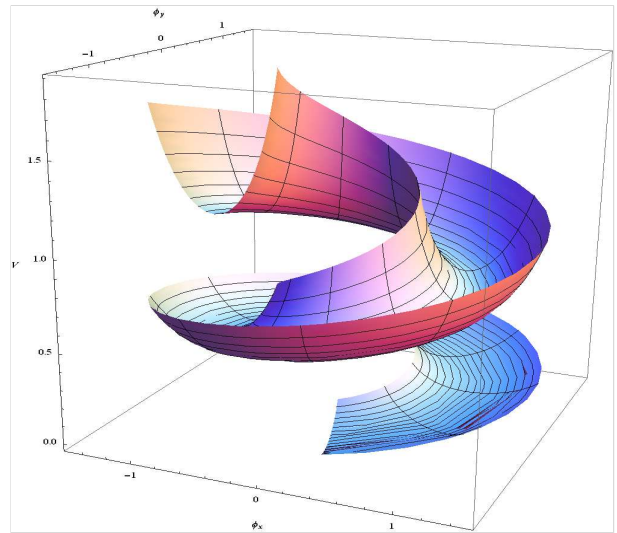}
\caption{The helicoid potential with unit $10^{-8} M_P^4$ . Along radial direction, the minimum of the potential locates at $|\Phi|\equiv r=1$, while the phase $\te$ provides a flat direction along the helix line, from which it is easy to get super-Planckian field excursion.}
\end{figure}\label{g1}

The potential (\ref{po}) is simple but actually has fancy helicoid structure, as shown in Fig. $1$.
The exponential factor $e^{r^2}$ does not depend on the phase $\te$ resulting from the global $U(1)$ symmetry of K\"ahler potential (\ref{ka}), consequently there is no $\eta$ problem for this phase inflation.
The complex field magnitude $|\Phi|\equiv r$ obtains vacuum expectation value at $\langle r\rangle=1$ as both $e^{r^2}\frac{1}{r^2}$ and $(\ln r)^2$ reach minimums at $r=1$.

\begin{figure}
\centering
\includegraphics[width=70mm, height=70mm,angle=0]{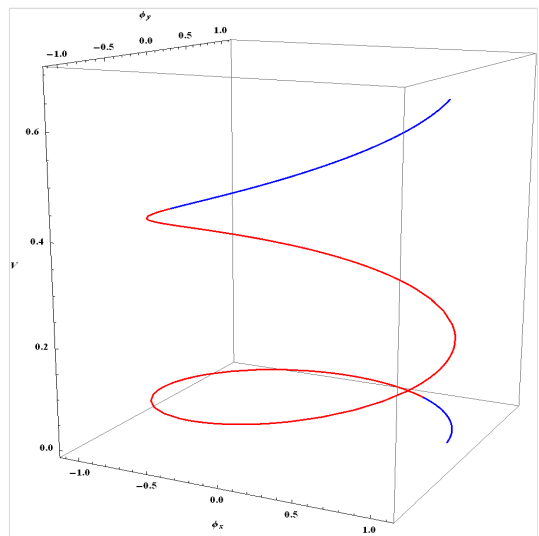}
\caption{Helix trajectory with $r=1$. The red part indicates the phase excursion for quadratic inflation with $N_e=55$.}
\end{figure}\label{g2}

The mass along the radial direction is
\begin{equation}
m_r^2=\frac{1}{2}\frac{\partial^2 V}{\partial r^2}|_{r=1}=(2+\frac{1}{\te^2})V_I, \label{mr}
\end{equation}
where the factor $\frac{1}{2}$ is from the normalization of $r$, and $V_I=ea^2\te^2$ is the potential for inflation.
Eq. (\ref{mr}) shows that the mass of $r$ is larger than the Hubble constant therefore the radial component is frozen out during inflation, and we realize the quadratic inflation dominated by $V_I$. The helical inflation path is shown in Fig. $2$.
The inflaton $\te$ has physical mass $m_\te=\sqrt{e}a$, at scale $10^{13}$ GeV from observations \cite{Ade:2013uln, Ade:2014xna}. It predicts the spectral index $n_s\simeq 1-\frac{2}{N_e}$
and the tensor-to-scalar ratio $r\simeq \frac{8}{N_e}$, where $N_e$ is the e-folding number.

In the PNGB inflation, the phase of the Higgs field also plays the role of the inflaton \cite{Freese:1990rb}. However, the potential is periodic and a super-Planckian decay constant is needed. In our model, the inflation path is helical, there is no limit on the field displacement during inflation, actually this is a new realization of the monodromy, which is proposed as stringy axion inflation in a rather different way \cite{McAllister:2008hb}.

The norm can be stabilized in the sub-Planckian scale by taking the following superpotential
\begin{equation}
W=a X \Phi^{-\frac{1}{n}}\ln\frac{\Phi}{\Lambda}.
\end{equation}
The scalar potential becomes
\begin{equation}
V=a^2e^{r^2}r^{-\frac{2}{n}}((\ln r-\ln\Lambda)^2+\te^2).
\end{equation}
The minimum of the factor $e^{r^2}r^{-\frac{2}{n}}$ locates at $r_0=\frac{1}{\sqrt{n}}(=\Lambda)$. The mass along the radial direction at $r_0$ is
\begin{equation}
m_r^2=(2+\frac{n}{\te^2})V_I>H^2,
\end{equation}
where $V_I=(en)^{1/n} a^2 \te^2$,
providing a strong stabilization even $r$ is very small.
Giving $n\geqslant 10$ the norm can be stabilized at $r_0\sim O(10^{-1})(M_P)$.

\subsection*{Monodromy from Supersymmetric Field Theory}
To realize helical phase inflation, the monodromy (\ref{mono}) of superpotential (\ref{sp}) is crucial. The monodromy is from the superpotential
\begin{equation}
W_0=\sigma X\Psi(T-\delta)+Y(e^{-\alpha T}-\beta \Psi)+Z(\Psi\Phi-\lambda), \label{sp1}
\end{equation}
in which the coupling constants of the last two terms are absorbed in the chiral superfields $Y$ and $Z$, and $\sigma\ll1$ to provide inflation potential at scale much lower than that of last two terms.
The couplings in (\ref{sp1}) consist of renormalizable perturbative terms  and $Ye^{-\alpha T}$, which is considered to be an effective description of certain non-perturbative effect. A reasonable decay constant $f$ is much smaller than Planck mass, so $\alpha\propto\frac{1}{f}\gg1$. In type \MyRoman{2} string theory similar non-perturbative term can be obtained from D-brane instanton effect \cite{Blumenhagen:2009qh}.

The supergravity vacuum is given by the vanishing F-term conditions
\begin{equation}
F_z=D_zW_0=\partial_z W_0+K_zW_0=0,
\end{equation}
where $z\in\{X, Y, Z, T, \Psi, \Phi\}$. Combing with Minkowski vacuum condition $W_0=0$, the preferred vacuum is given by $\partial_zW_0=0$, and it locates at
\begin{equation}
\begin{split}
&\langle X\rangle=\langle Y\rangle=\langle Z\rangle=0, ~\langle T\rangle=\delta, \\
&\langle \Psi\rangle=\frac{1}{\beta}e^{-\alpha \delta},
\langle \Phi\rangle=\lambda \beta e^{\alpha \delta}.
\end{split}
\end{equation}
Giving $\langle \Phi\rangle\gg \langle \Psi\rangle$, near the vacuum $Y, Z, T, \Psi$ obtain heavy effective masses from the last two coupling terms while $X, \Phi$ are light. During inflation all the heavy fields are frozen out and can be integrating out, then we get an effective field theory at inflation scale. To integrate out the heavy fields, we need to solve the equations of vanishing F-terms of frozen fields
\begin{equation}
\begin{split}
&F_Y=e^{-\alpha T}-\beta \Psi+K_YW_0=0, \\
&F_Z=\Psi\Phi-\lambda+K_ZW_0=0.
\end{split} \label{YZ}
\end{equation}
In minimal supergravity, the K\"ahler potential is $K=\Sigma z\bar{z}$ \footnote[3]{Except $T$, for the reasons shown below, K\"ahler potential of $T$ has to be shift invariant under $T\rightarrow T+iC$.}, which gives $K_z=\bar{z}$. Besides,
near the vacuum $Y= Z\approx0\ll M_P$, the higher order terms $K_zW_0$ in (\ref{YZ}) just give small corrections and we get the approximate solutions of Eq.~(\ref{YZ})
\begin{equation}
\Psi=\frac{\lambda}{\Phi}, ~~~~T=\frac{1}{\alpha}\ln \frac{\Phi}{\beta\lambda}.
\end{equation}
Substituting above solutions for $T$ and $\Psi$ in the original superpotential $W_0$, we get the effective superpotential (\ref{sp}) during inflation.
The parameters should satisfy
\begin{equation}
\beta e^{\alpha\delta}=\lambda^{-1}\gg1,~~ a=\frac{\sigma\lambda}{\alpha}\sim10^{-5},
\end{equation}
which can be easily adjusted to fit with observations.

The singularity of superpotential $W$ at $\Phi=0$ is clear from this procedure. When $\Phi\rightarrow0$, $\Psi\gg\Phi$ constrained by (\ref{YZ}) and it is illegal to integrate out $\Psi$, the model should be studied in another effective field theory. Fortunately during inflation $|\Phi|$ is fixed at VEV and the phase rotation cannot break the effectiveness of the theory given by $W$.
As to the monodromy, vanishing conditions of $F_Y$ and $F_Z$ fix four directions of three complex fields $T, \Psi$ and $\Phi$, but allow the transformation
\begin{equation}
\begin{split}
&\Psi\rightarrow\Psi e^{-u-iv} \\
&\Phi\rightarrow\Phi e^{u+iv}  \\
&T\rightarrow T+u/\alpha+iv/\alpha.
\end{split} \label{tr}
\end{equation}
However, because of the supergravity correction on the scalar potential $V\propto e^K$, norms of $\Psi$ and $\Phi$ are stabilized, $u=0$ in (\ref{tr}). Field stabilization does not fix the phase rotation\footnote[4]{If the K\"ahler potential of $T$ is minimal, then the supergravity correction $e^{T\bar{T}}$ would fix the phase rotation as well!}, and for a whole circular rotation, $W_0\rightarrow W_0+2\pi \sigma iX\Psi/\alpha$, which is exact the monodromy in (\ref{mono}).

By integrating out the heavy fields, the supergravity correction $e^K$ should be replaced by the solution of Eq.~(\ref{YZ}) as well, which just gives norm-dependent terms and slightly shifts the fixed norm since $K$ is invariant under phase rotation. Specifically, for $T$ a shift symmetry in $K$ is needed, otherwise the factor $e^K$ contains phase of $\Phi$ and breaks the inflation. Among these phase factors, the phase of $\Phi$, after canonical field redefinition, has lightest physical mass and evolves as inflaton.

At quantum level, because of the non-renormalization theorem for the superpotenital, the loop corrections from integrating out heavy fields appear in K\"ahler potential only, and these corrections just sightly affect the field stabilization but not the phase inflation
which is protected by the $U(1)$ symmetry in K\"ahler potential.

\subsection*{UV Sensitivity of Large Field Inflation}
The crucial challenge for large field inflation is the higher dimensional operators from quantum gravity corrections \cite{Baumann:2014nda}. The higher order terms of the inflaton $\phi$
\begin{equation}
\Delta V=c_iV(\frac{\phi}{M_P})^{i}+\cdots, \label{co}
\end{equation}
are unignorable at the initial stage of inflation when $\phi\sim O(10)M_P$. They can modify the predictions significantly or even destroy slow-roll conditions. In this model, the inflaton is just the phase of a complex field like PNGB and admits no polynomial correction at all, in consequence quantum gravity corrections like (\ref{co}) immediately disappear without any constraint from extra-symmetry. So the helical phase inflation is not sensitive to the quantum gravity effects.

In the bottom-up approach, one can apply axionic shift symmetry of the inflaton $\phi\rightarrow\phi+c$, which is broken down to discrete symmetry $\phi\rightarrow \phi+2\pi f$ by non-perturbative effect. To fit the experimental observations it requires super-Planckian axion decay costant $f$ \footnote[5]{Giving a coupling between the inflaton kinetic term and Einstein tensor, natural inflation with $f\ll M_p$ still works \cite{Germani:2010hd}.},
which can be realized by aligned axions \cite{Kim:2004rp} (or equally a $S_2$ symmetry between two K\"ahler moduli \cite{Li:2014lpa}) or anomalous
gauged $U(1)_X$ with large gauge symmetry \cite{Li:2014xna}. The inflation path of aligned axions has similar helical structure in axion space \cite{Choi:2014rja, Tye:2014tja}, and it shows that the alignment mechanism is kind of monodromy inflation realized by axions that are plentiful in string compactification. Stringy inflation is expected to solve the UV sensitivity of large field inflation but needs to address several difficult problems like moduli stabilization, Minkowski or de Sitter vacua, etc. Our model provides another type of monodromy inflation just in supersymmetric field theory, which is more simpler and controllable. The $U(1)$ symmetry is build-in the K\"ahler potential and there is no naturalness problem in the top-down perspective.
Based on the supersymmetric field realization of inflation, a unified description of the inflation and the well-known GUT is at hand. A direct test on the relationship between inflation and GUT is the reheating process. In our model, a simple guess is the chiral superfield $X$ is a gauge singlet in certain grand unification model, like the scenario in \cite{Harigaya:2014roa}, then the inflaton decays into visible particles through couplings of $X$ during reheating.

\subsection*{Conclusion}
We have shown in this letter that the phase inflation along a {\it{single} helix} trajectory can be realized in a surprisingly simple way based on minimal supergravity. The global $U(1)$ symmetry of minimal K\"ahler potential naturally solves the $\eta$ problem which appears generically for supergravity inflation. The radial direction is strongly stabilized during inflation, and the super-Planckian phase excursion is fulfilled along a helix path.

The helical phase inflation is not sensitive to the quantum gravity effect as higher order corrections are not possible for a PNGB like particle. The phase inflation also admits an effective description on super-Planckian field excursion within supersymmetric field theory, and it naturally leads to field monodromy, which relates to a global $U(1)$ symmetry explicitly breaking at inflation scale. It is surprising that {\it the supergravity $\eta$ problem, field stabilization, puzzle of super-Planckian field excursion and monodromy inflation admit a simple unified solution within a helicoid structure}. As will be shown in \cite{Li:2014dec}, the monodromy in (\ref{sp1}) can be easily generalized to obtain supersymmetric field realization of aligned axions with consistent field stabilization \cite{Kim:2004rp, Choi:2014rja, Tye:2014tja}, so the helical phase inflation actually provides a general frame to realize supergravity inflation with several amazing features. However, because inflation is an extraordinary unusual and unique event in the history of our Universe, we are not hesitant in being bold.  It will be phenomenal if nature employed helix structures to promote evolution from the very early universe to present time organisms.

\noindent {\bf Acknowledgements. }
DVN would like to thank Andriana Paraskevopoulou
for inspiration and discussions during the writing of this paper.
The work of DVN was supported in part
by the DOE grant DE-FG03-95-ER-40917. The work of TL is supported in part by
    by the Natural Science
Foundation of China under grant numbers 10821504, 11075194, 11135003, 11275246, and 11475238, and by the National
Basic Research Program of China (973 Program) under grant number 2010CB833000.

\end{document}